\documentclass[article,letterpaper,prc,flaotfix,nofootinbib,aps]{revtex4}

\usepackage{amsmath}
\usepackage{amsfonts}
\usepackage{amssymb}
\usepackage[english]{babel}
\usepackage{graphicx}
\usepackage{color} 
\usepackage{dcolumn} 

\newcommand {\vn}	{$v_2$}

\newcommand {\pt}	{$p_T$}
\newcommand {\pTRef}	{$p_T^{\rm{Ref}}$}
\newcommand {\pp}	{\textit{p+p}}
\newcommand {\sNN}	{$\sqrt{s_{\rm{NN}}} = 200$ GeV}

\newcommand {\dphi}	{\Delta\phi}
\newcommand {\cij}	{\cos2\dphi_{}}

\newcommand {\vtwo}[1]	{v_{2}}
\newcommand {\vv}[1]	{v_2}
\newcommand {\vsq}[1]	{v_2^2}
\newcommand {\mean}[1]	{\langle#1\rangle}
\newcommand {\dd}[1]	{\delta_{2} }
\newcommand {\dsq}[1]	{\delta_2^{2} }
\newcommand {\Vdn}		{V_{n\Delta} }
\newcommand {\Vdtwo}		{V_{2\Delta} }
\newcommand {\vnDiag}		{v_{n}^{\rm Diag} }
\newcommand {\vTwoDiag}		{v_{2}^{\rm Diag} }
\newcommand {\dTwoDiag}		{\delta_{2}^{\rm Diag} }
\newcommand {\dTwo}	{\delta_{2} }

\newcommand {\acl}	{{d}}

\newcommand {\Nhy}	{N_{hy}}
\newcommand {\Ncl}	{N_{cl}}
\newcommand {\Na}	{N_{\acl}}

\begin{document}

\author{Daniel Kiko\l{}a$^1$, Li Yi$^1$, ShinIchi Esumi$^2$, Fuqiang Wang$^1$, Wei Xie$^1$}
\affiliation{$^1$Department of Physics, Purdue University, West Lafayette, IN 47907, USA \\
$^2$Institute of Physics, University of Tsukuba, Tenno-dai 1-1-1, Tsukuba, Ibaraki 305-8571, Japan
}

\title{Nonflow ``factorization'' and a novel method to disentangle anisotropic flow and nonflow}

\date{\today}

\begin{abstract}

Anisotropic flow, which is sensitive to hydrodynamic expansion of the medium created in heavy ion collisions, has been extensively studied. Anisotropic flow measured using azimuthal correlations of final state particles is therefore contaminated by correlations unrelated to the reaction plane (nonflow). Currently anisotropic flow and nonflow cannot be separated experimentally in a model independent manner. Using PYTHIA simulations of \pp\ collisions we show that nonflow  approximately factorizes in transverse momentum. This fact may be used to disentangle flow and nonflow in heavy ion data by performing a two-component factorization fit to Fourier coefficients of two-particle cumulants.

\end{abstract}

\maketitle

\section{Introduction}

The azimuthal distribution of particle momenta is an important observable in studying properties of nuclear matter created in relativistic heavy ion collisions. Pressure driven hydrodynamic expansion of the created medium transforms the initial configuration anisotropy of the colliding nuclei into the final state momentum anisotropy~\cite{Ollitrault:PhysRevD.46.229}. The initial spatial anisotropy decreases rapidly with the system expansion. Therefore the final state azimuthal anisotropy is primarily sensitive to the early stages of the system evolution \cite{Kolb:2003dz,PhysRevC.82.054904}.

Particle distributions are often quantified by the Fourier series in azimuthal angle $\phi$ relative to the symmetry plane $ \Psi_n$:
\begin{equation}
\frac{{\rm d}^2 N}{{\rm d} p_T {\rm d}\phi} \propto 1 + \sum_{n=1}^{\infty} 2v_n (p_T) \cos(n(\phi- \Psi_n )) \, .
\end{equation}
The symmetry plane $\Psi_n$ is often called participant plane. It can be different from the reaction plane (which is defined by beam direction and impact parameter vector), and can be different for different $n$ due to fluctuations~\cite{Flow:fluctuations}. Fourier coefficients $v_n = \langle \cos(n(\phi- \Psi_n )) \rangle$ characterize the anisotropy for different harmonics. The second coefficient, \vn, is called elliptic flow. It has been extensively studied and has provided compelling evidence for the creation of the strongly interacting quark gluon plasma at RHIC~\cite{STAR:white:paper,BRAHMS:white:paper,PHENIX:white:paper,PHOBOS:white:paper}. Studies of $v_n$ of all harmonics are expected to yield valuable information about the properties of the quark gluon plasma \cite{Lacey:2010hw,Flow:fluctuations,Schenke:2010rr,Kolb:2003zi}. 

The participant plane is experimentally unknown and is estimated by the event plane $\Psi_{\rm{ EP}}$ which is constructed using all charged particles in an event~\cite{PhysRevC.58.1671}. Because of this, intrinsic particle correlations unrelated to the participant plane may contribute to the anisotropy. This contribution is generally called nonflow. Nonflow correlations include, for instance, jet correlations and resonance decays.  

Anisotropic flow can be also measured using the two particle correlation method. It was shown that the event plane and two-particle correlation methods are approximately equivalent~\cite{Trainor:2008jp}. The two particle azimuthal correlation function can be expanded in a Fourier series in relative angle $\Delta \phi$ of the particle pair a,b:
\begin{equation}
\frac{{\rm d}^3N^{\rm Pair}}{ {\rm d} p_T^a {\rm d} p_T^b {\rm d} \Delta \phi} \propto 1 + \sum_{n=1}^{\infty} 2 \Vdn (p_T^a,p_T^b) \cos(n\Delta \phi)
\end{equation}
In the case where correlations are only due to flow, particles are independently correlated to the participant plane, so $\Vdn$ should factorize \cite{Borghini:PhysRevC.62.034902}:
\begin{equation}
\Vdn(p_T^a,p_T^b) = v_{n}(p_T^a) v_{n}(p_T^b)	\; .
\label{Eq:FlowFactorization}
\end{equation}
Thus $v_n$ can be obtained for a given \pt\ by
\begin{equation}
 \vnDiag(p_T)  = \sqrt{\Vdn(p_T,p_T)}		\; .
 \label{Eq:v2Diagonal}
\end{equation}
Alternatively, if a reference $v_n(p_T^{\rm Ref})$ is known, $v_{n}$ can be obtained by mixed-\pt\ pair:
\begin{equation}
 v_{n}(p_T)  = \frac{\Vdn(p_T,p_T^{\rm Ref})}{v_n(p_T^{\rm Ref})}	\; .
  \label{Eq:v2OffDiagonal}
\end{equation}
In this paper we use $v_n$ to stand for anisotropies measured by two-particle correlations ($v_n\{2\}$ as often used in literature \cite{PhysRevC.64.054901}).

The flow factorization is a consequence of global correlations in an event: particles are all correlated to the common symmetry plane and particles are independent of each other. Nonflow, on the other hand, due to intrinsic correlations of particles, would not factorize a priori. Therefore two particle correlations in heavy ion events, including both flow and nonflow, would not generally factorize. However, a global fit to two-particle azimuthal correlations in Pb+Pb collisions using Eq. (\ref{Eq:FlowFactorization}) yielded satisfactory results at low-\pt~\cite{Alice:FlowFactorization}. This led to the suggestion that low-\pt\ correlations are dominated by anisotropic flow. 

In this paper we will show, by using PYTHIA simulations, that nonflow approximately factorizes, contrary to common perception. This is not surprising because jet fragments are all correlated to the jet axis, $\Psi_{\rm jet}$, analogues to the symmetry plane in heavy-ion event. In the case of jet fragmentation, there is a conserved total momentum of thrust; analogously for the case of anisotropic flow, there is conserved total transverse momentum of zero. Except for this common correlation, jet fragments may otherwise be independent on each other. This would result in factorization  of two-particle  correlation in a jet: $\mean {\cos (\phi_1 - \phi_2)} = \mean {\cos (\phi_1 - \Psi_{\rm jet})}\mean {\cos (\phi_2 - \Psi_{\rm jet})}$.  This suggests that, although flow factorizes, factorization does not necessarily imply flow. In other words, the observed factorization in data alone does not necessarily lead to the conclusion that the correlations are dominated by flow.

In the second part of the paper we will explore the possibility to extract flow and nonflow using a two-component factorization fit to mixed events of PYTHIA \pp\ collisions embedded in toy-model Monte Carlo simulations of hydrodynamical flow.

\section{Nonflow factorization \label{Sec:Nonflow:factorization}}

One of the main sources of nonflow correlations in \pp\ collisions is jets. Particles in a jet are produced by fragmentation of a single high-\pt\ parton, and are therefore correlated with the direction of the parton (jet axis). High-\pt\ particles in a jet tend to be aligned with the jet axis while low-\pt\ particles have a lesser degree of collinearity relative to the jet direction. Effectively, fragmentation process produces a ``global'' correlation with respect to the jet axis, and the correlation depends on \pt. If jet fragments are indenpendent of one another, then nonflow (jet correlations) could also factorize.  

We investigate nonflow factorization using 210 M  \pp\  events at $\sqrt{s} = 200$ GeV generated using PYTHIA 6.4~\cite{Pythia} with default parameters except that we require a minimum momentum transfer of 3 GeV to ensure jet production in our simulations. Only stable charged particles are used in our study (pions, kaons, protons and antiprotons). The \pt\ spectrum is shown in Fig. \ref{Fig:pTspectrum:v2VsPtMixed} (a). 

In  the experimental analysis of two-particle correlations an $\eta$ gap often is applied to reduce nonflow from resonance decays and intra (near-side) jet correlations. However inter (away-side) jet correlations are still present. We require $\Delta \eta > 1$ in our study: we correlate each particle from the backward pseudorapidity region ($- 1 < \eta < - 0.5$) with particles at forward pseudorapidity ($0.5 < \eta < 1$). Two-particle correlation Fourier coefficients $\Vdn$ are calculated using the direct Q-Cumulants method \cite{Bilandzic:2010jr,Yi:2011hs}, which avoids nested loops in the analysis. The Q-vector for a given \pt\ bin is defined by: 
\begin{equation}
Q_{n}(p_T) = \sum_{i=1}^{N} e^{in \phi_i}	
\end{equation}
where $N$ is the particle multiplicity used in analysis. For two exclusive regions the two particle azimuthal moment is given by $Q_{n}(p_T^a)Q_{n}^{*}(p_T^b)/(N(p_T^a)N(p_T^b))$ where $N(p_T^a)$, $N(p_T^b)$ are multiplicities in those two $\eta$ regions. The two particle coefficient $\Vdn$ is the event average:
\begin{equation}
\Vdn(p_T^a,p_T^b) =  \left \langle \frac{Q_{n}(p_T^a)Q_{n}^{*}(p_T^b)}{N(p_T^a)N(p_T^b)} \right \rangle \, .
\end{equation}
The error on $\Vdn$ is simply $\sigma/\sqrt{N_{\rm evt}}$, where $\sigma$ is sample standard deviation of $\Vdn$ and $N_{\rm evt}$ is number of events. We treat all errors as uncorrelated.

\begin{figure}[htdp]
\begin{center}
\begin{tabular}{ccc}
\includegraphics[width=0.32\textwidth]{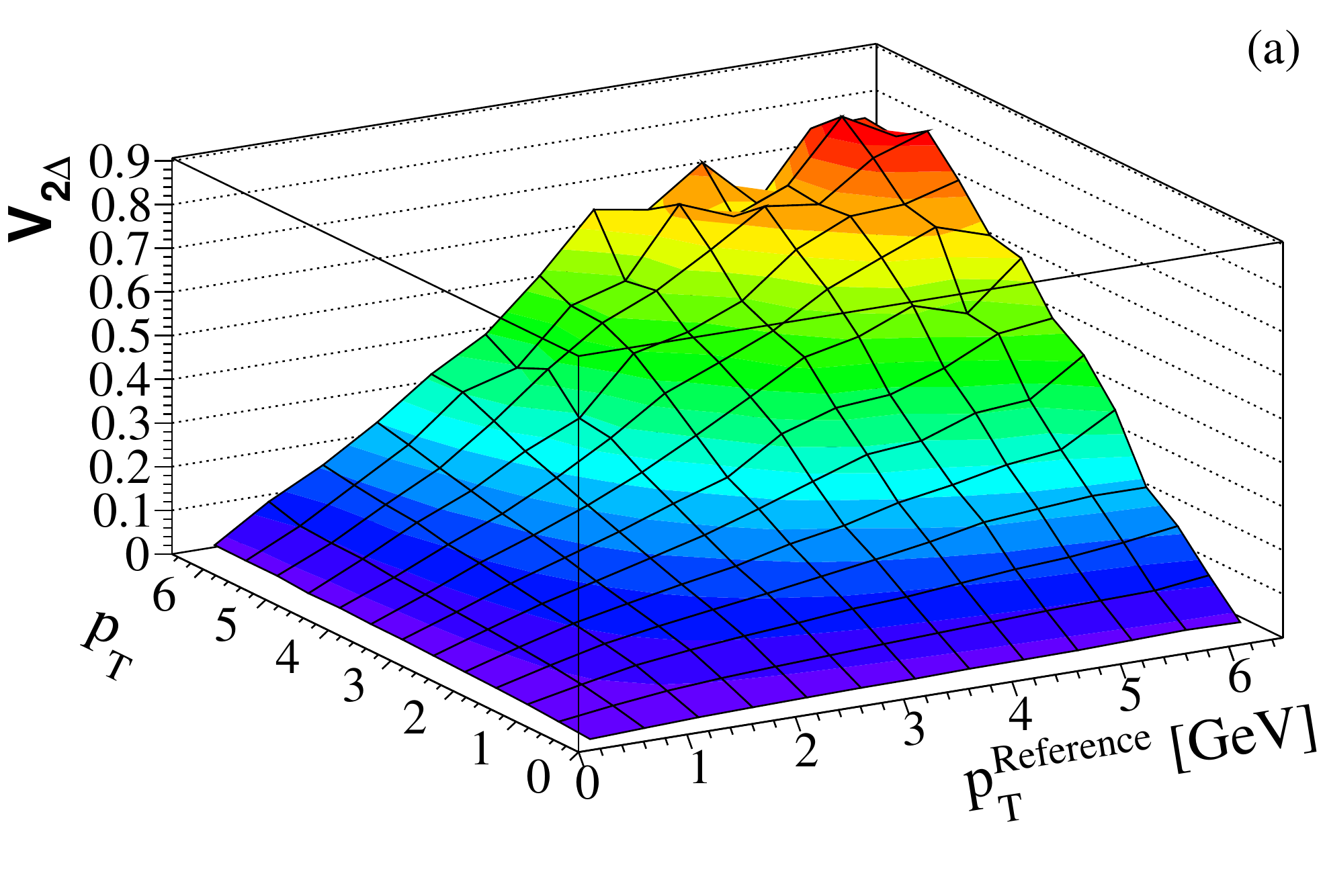} &
\includegraphics[width=0.32\textwidth]{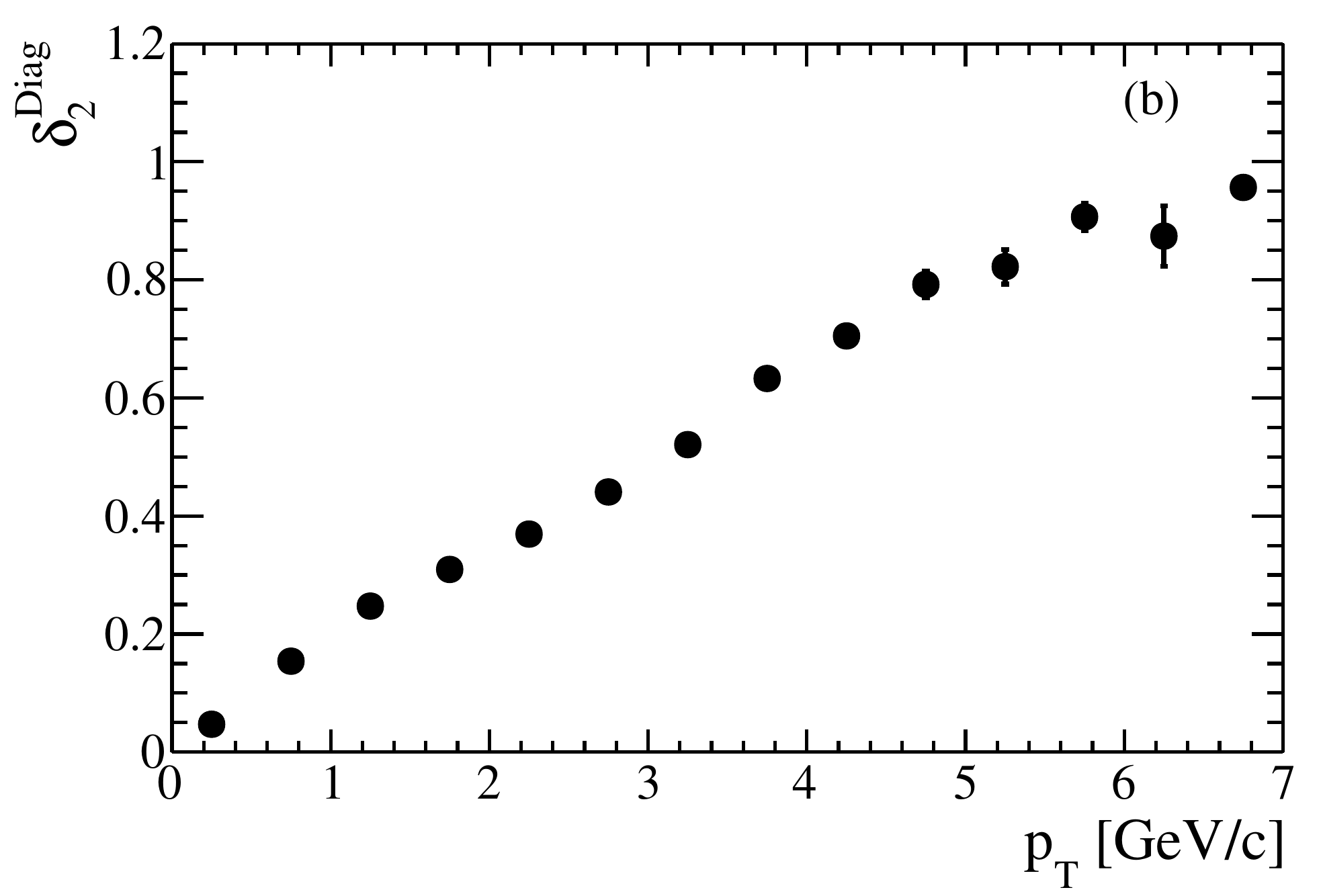} &
\includegraphics[width=0.32\textwidth]{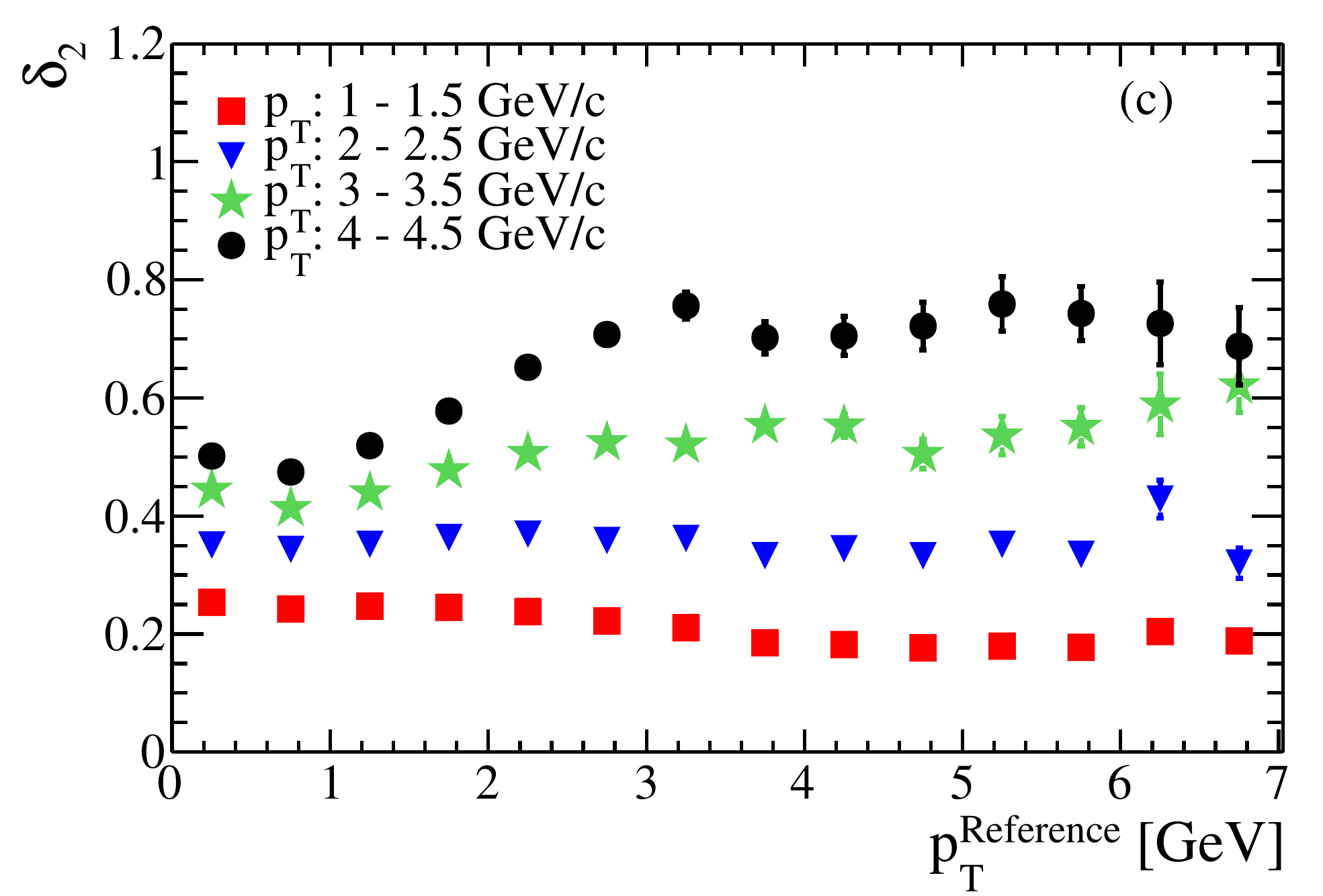} \\
\end{tabular}
\caption{\label{Fig:V2vsPtRef}(Color online) (a) $\Vdtwo(p_T,p_T^{\rm{Ref}})$ distribution in PYTHIA \pp. (b) $\dTwoDiag$ as a function of \pt,  (c) $\delta_{2}$ as a function of \pTRef\ for several \pt\ bins.} 
\end{center}
\end{figure}

Figure \ref{Fig:V2vsPtRef}(a) shows the two particle Fourier coefficient $\Vdtwo$ obtained for PYTHIA \pp\ events. The correlation increases with \pt\ as expected as the particles are more closely aligned along the jet axis. Now we test applicability of the factorization hypothesis to correlation in \pp\ events. First, we calculate the single Fourier coefficient for nonflow, $\dTwo$, from the diagonal points of $\Vdtwo$ distribution by Eq. (\ref{Eq:v2Diagonal}). This is presented in Fig. \ref{Fig:V2vsPtRef}(b). If factorization holds we expect that $\dTwo$ obtained from $\Vdtwo$ of mixed-\pt\ pairs (\pt,\pTRef) by Eq. (\ref{Eq:v2OffDiagonal}) to be independent of \pTRef. Figure \ref{Fig:V2vsPtRef}(c) shows single Fourier component $\dTwo$(\pt) as a function of \pTRef. In the case of lower \pt\ $(1.5 - 2$, $2-2.5$), $\dTwo$ is approximately constant in the whole considered range of \pTRef, while for the high \pt\ ($3-3.5$ and $4 -4.5$ GeV/c) $\dTwo(p_T) \approx \rm{const}$ at $p_T^{\rm{Ref}}> 3$~GeV/c. These results suggest that the nonflow correlations may approximately factorize in a limited \pt\ range. 

\begin{figure}[htdp]
\begin{center}
\begin{tabular}{cc}
\includegraphics[width=0.45\textwidth]{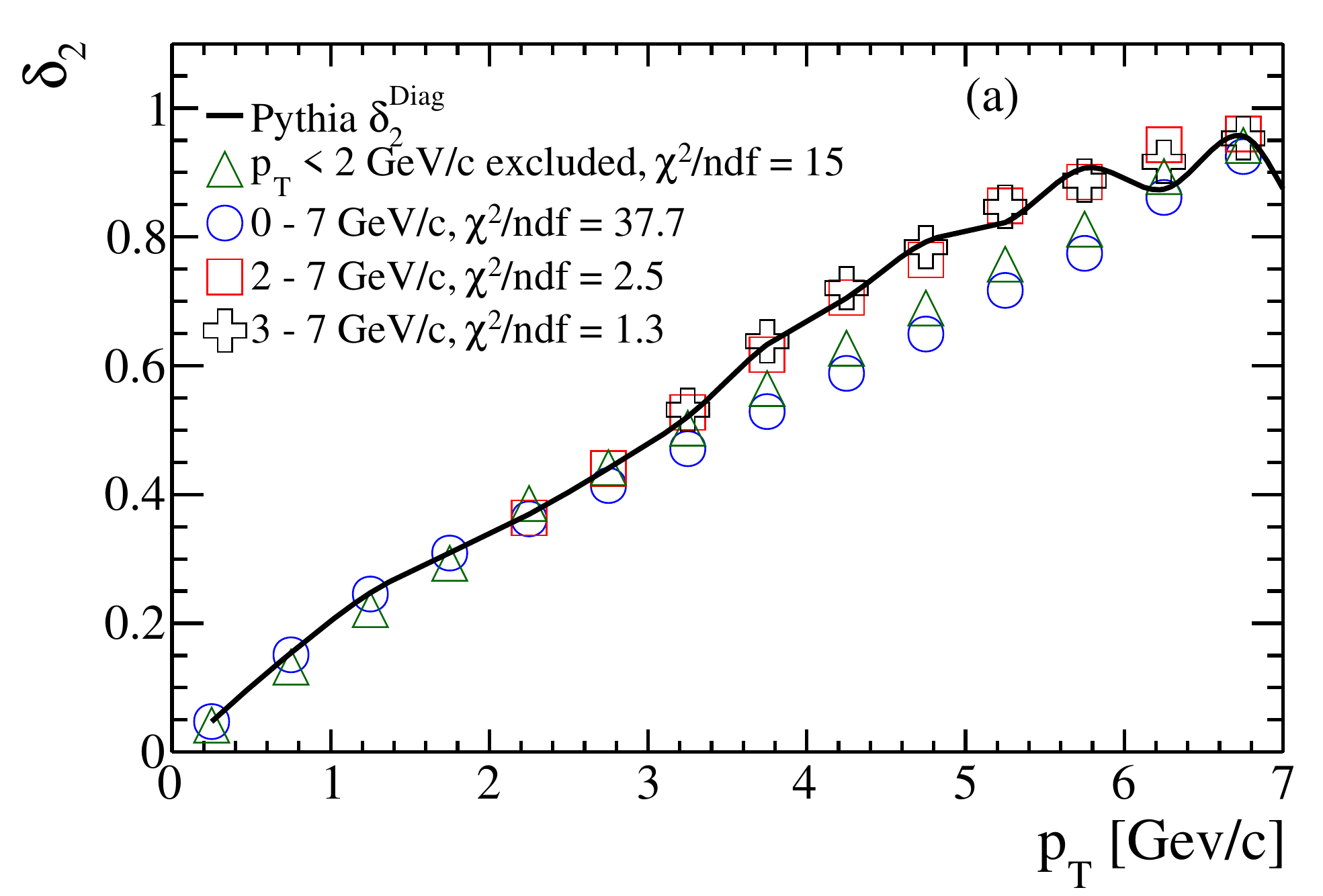} &
\includegraphics[width=0.45\textwidth]{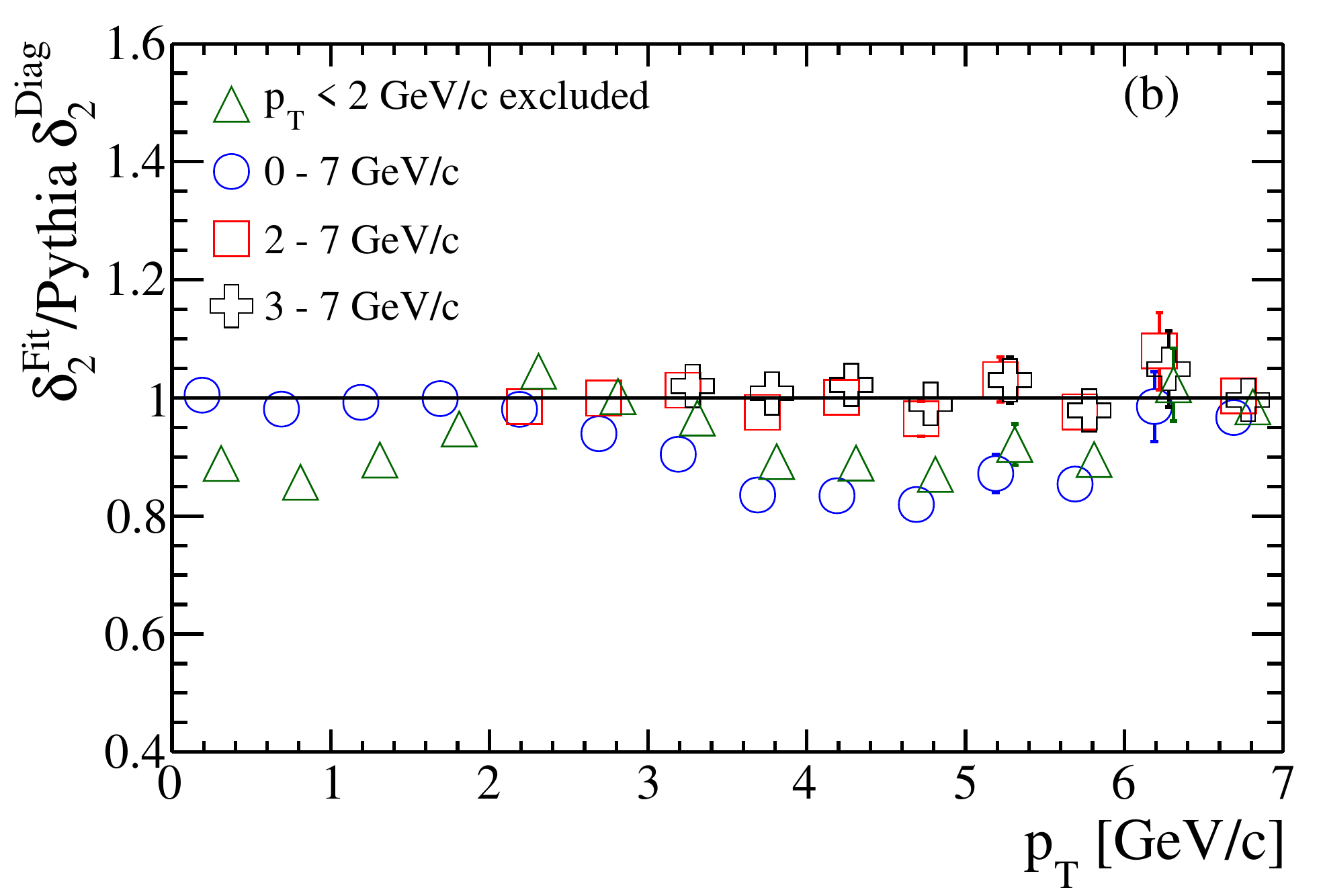} \\
\end{tabular}
\caption{\label{Fig:VnVnFitsPythiaSingleComponent}(Color online) (a) $\dTwo(p_T)$ from Eq. (\ref{Eq:FlowFactorization}) fit to PYTHIA \pp\ simulations in various \pt\ ranges compared to $\dTwoDiag$ (see text for details). (b) Ratio of fit results to $\dTwoDiag$.} 
\end{center}
\end{figure}

In order to obtain a quantitative description of applicability of nonflow factorization, we fit $\Vdtwo(p_T^a,p_T^b)$ data of PYTHIA \pp\ events with Eq. (\ref{Eq:FlowFactorization}). Figure \ref{Fig:VnVnFitsPythiaSingleComponent} shows the results. Because we expect approximate factorization to work best at high-\pt, we compare fit results in different ranges of $p_T^a,p_T^b$ pairs. We examine the fit quality for all \pt\ pairs (open circles in Fig. \ref{Fig:VnVnFitsPythiaSingleComponent}),  excluding pairs with the lowest \pt\ only i.e. $\Vdtwo(p_T^a,p_T^b)$ points are excluded from fit if $p_T^a<p_T^{\rm{min}}$ and $p_T^b <p_T^{\rm{min}}$ (open triangles in Fig. \ref{Fig:VnVnFitsPythiaSingleComponent}), and using only pairs with high-\pt\ i.e. $p_T^a >p_T^{\rm{min}}$ and $p_T^b >p_T^{\rm{min}}$ (open square and crosses in Fig. \ref{Fig:VnVnFitsPythiaSingleComponent}). The goodness of the fit is assessed with $\chi^2/{\rm{NDF}}$, where NDF is number of degree of freedom. In most cases the fit is driven by low-\pt\ points and Eq. (\ref{Eq:FlowFactorization}) underestimates $\Vdtwo$ at high \pt. This may suggest that at low \pt\  there are additional processes that are not correlated with jet production. Fit quality ($\chi^2/{\rm NDF}$) improves when a higher threshold $p_T^{\rm{min}}$ is applied, and the results are in good agreement with numerical data at high-\pt\ ($p_T>3$~GeV/c). Good fit quality at high-\pt\ is a convincing argument that nonflow due to jet fragmentation factorizes at higher \pt. 

\section{Novel means to disentangle flow and nonflow in Heavy Ion collisions}

The relative contributions of flow and nonflow to the measured final state momentum anisotropy is an important question and several methods were proposed to distinguish them. Ollitrault \textit{et al.} suggested that nonflow could be estimated using measurements in \pp\ collisions \cite{Ollitrault:PhysRevC.80.014904}, with the assumption that nonflow in heavy-ion collisions is identical to correlations in elementary \pp\ collisions.
The PHOBOS experiment improved this approach by measuring azimuthal anisotropy using particles with a large $\eta$ gap, estimating long-range nonflow contribution using \pp\ data and Monte Carlo event generators \cite{PHOBOS:PhysRevC.81.034915}. Another approach was proposed in Ref. \cite{Daugherity:2008su} where the measured correlation raw data were decomposed into elliptic flow and minijet correlations, where the minijet component was assumed to consist of a two-dimensional Gaussian in ($\Delta \eta, \Delta \phi$) and 
a dipole moment in $\Delta \phi$. In Ref. \cite{Yi:2011hs}, it was shown that if flow fluctuations are Gaussian, then it may be possible to disentangle flow, flow fluctuations and nonflow by simultaneous measurement of two-, four- and six-particle azimuthal cumulants. 

The approximate factorization of nonflow may provide us with a new means to  separate flow and nonflow in heavy ion collisions and obtain their \pt\ dependence. In a cluster model \cite{Nonflow:cluster:model} where particles are composed of hydro particles and correlated particles from clusters, the final two-particle anisotropy for a given \pt\ bin is 
\begin{equation}
\Vdtwo(p_T,p_T)=\left(\frac{\Nhy}{N}v_{2,hy}+\sum_{cl}\frac{\Ncl\Na}{N}v_{2,\acl}\right)^2+\sum_{cl}\frac{\Ncl\Na^2}{N^2}\left(\mean{\cij}_{cl}\right) \; .
\label{Eq:GenaralV22}
\end{equation}
Here $\Nhy$ is the number of hydro-medium particles, $v_{2,hy}$ - elliptic flow of hydro-medium particles, $\Ncl$ - average number of clusters of one type, $\Na$ - number of daughter particles in a cluster, $\sum_{cl}$ runs over all types of clusters, and $v_{2,\acl}$ is elliptic flow of cluster daughter particles which acquire flow due to anisotropy of clusters; $\mean{\cij}_{cl}$ is the average cosine of twice pair opening angle in the cluster and is nonflow. Note that Eq.\ref{Eq:GenaralV22} slightly differs from Eq.~22 in Ref.~\cite{Nonflow:cluster:model} because we have assumed a Poisson distribution in number of clusters whereas in Ref.~\cite{Nonflow:cluster:model} the number of clusters was fixed. For a pair $\left ( p_T^a, p_{T}^b \right )$, Eq. (\ref{Eq:GenaralV22}) can be generalized into 
\begin{equation}
\Vdtwo(p_T^a,p_T^b) = v_2(p_T^a) v_{2}(p_T^b) + \sum_{cl}\frac{\Ncl\Na(p_T^a) \Na(p_T^b) }{N(p_T^a) N(p_T^b) }\mean{\cij}_{cl} \\
\label{Eq:GenaralV22:PairPt}
\end{equation}
where \begin{equation}
v_2(p_T) =  \frac{\Nhy}{N}v_{2,hy}+\sum_{cl}\frac{\Ncl\Na}{N}v_{2,\acl}
\label{Eq:Flow:ClusterModel}
\end{equation}
represents the overall anisotropic flow. If the nonflow term factorizes and only one type of clusters dominate,
\begin{eqnarray}
\frac{\Ncl\Na(p_T^a) \Na(p_T^b) }{N(p_T^a) N(p_T^b) }\mean{\cij}_{cl}  &=& \left( \sqrt{\Ncl} \frac{\Na(p_T^a)}{N(p_T^a)}\mean{\cos 2 \left( \phi^a - \Psi_{cl} \right) } \right) 
\left( \sqrt{\Ncl} \frac{\Na(p_T^b)}{N(p_T^b)}\mean{\cos 2 \left( \phi^b - \Psi_{cl} \right) } \right) \; , \nonumber \\ 
& \equiv & \delta_2(p_T^a) \delta_2(p_T^b) \; ,
\end{eqnarray}
then $\Vdtwo$ can be expressed as sum of two components, each of which factorizes:
\begin{equation}
\Vdtwo(p_T^a,p_T^b) = v_{2}(p_T^a)v_{2}(p_T^b) + \delta_2(p_T^a) \delta_2(p_T^b)	\; .
\label{Eq:FlowNonFlowFactorization}
\end{equation}
Here $\delta_2$ is the ``single particle'' nonflow,
\begin{equation}
\delta_2 =   \sqrt{\Ncl} \frac{\Na}{N}\mean{\cos 2 \left( \phi - \Psi_{cl} \right) }	\; ,
\label{Eq:NonFlow:ClusterModel}
\end{equation}
where $\Psi_{cl}$ is a cluster axis. Note that in the literature~\cite{PhysRevC.58.1671} the symbol $\delta$ is often used to stand for pair-wise nonflow, i.e. the second term of Eq. (\ref{Eq:GenaralV22:PairPt}) or (\ref{Eq:FlowNonFlowFactorization}). In this paper we use $\delta$ to stand for factorized ``single particle'' nonflow, and the pair-wise nonflow is $\delta_2(p_T^a) \delta_2(p_T^b)$. 

\begin{figure}[htdp]
\begin{center}
\begin{tabular}{cc}
\includegraphics[width=0.45\textwidth]{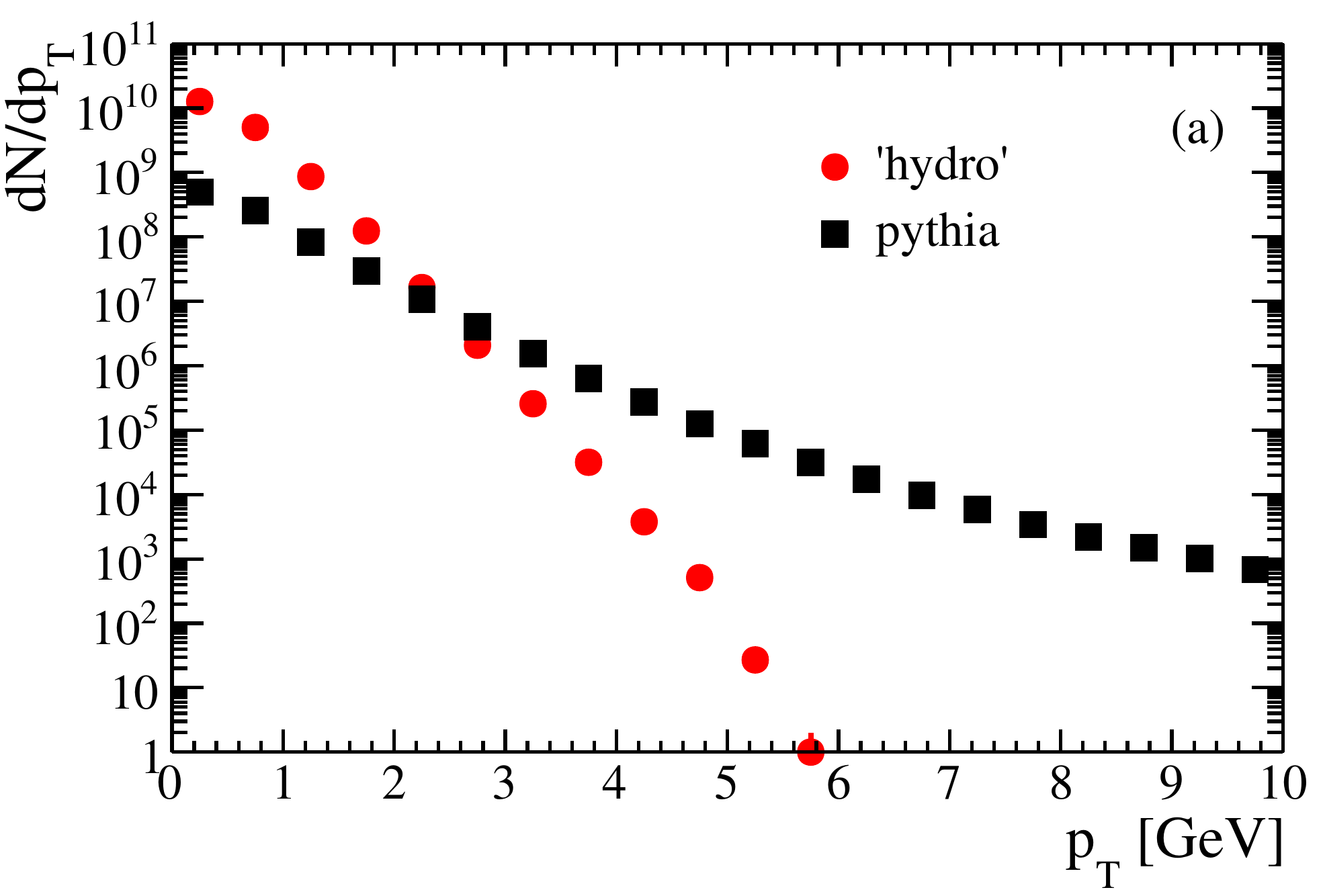}  &
\includegraphics[width=0.45\textwidth]{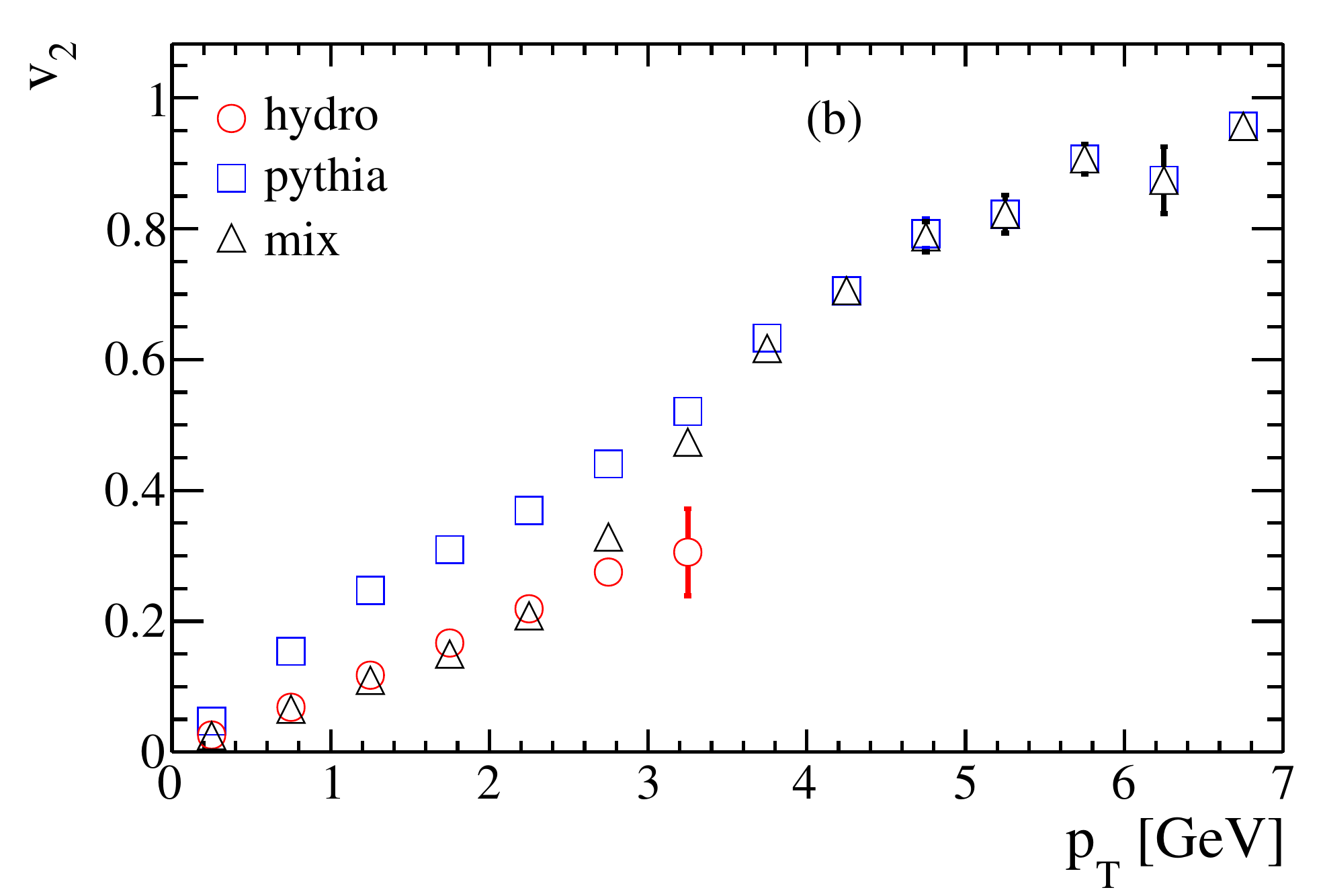} \\
\includegraphics[width=0.45\textwidth]{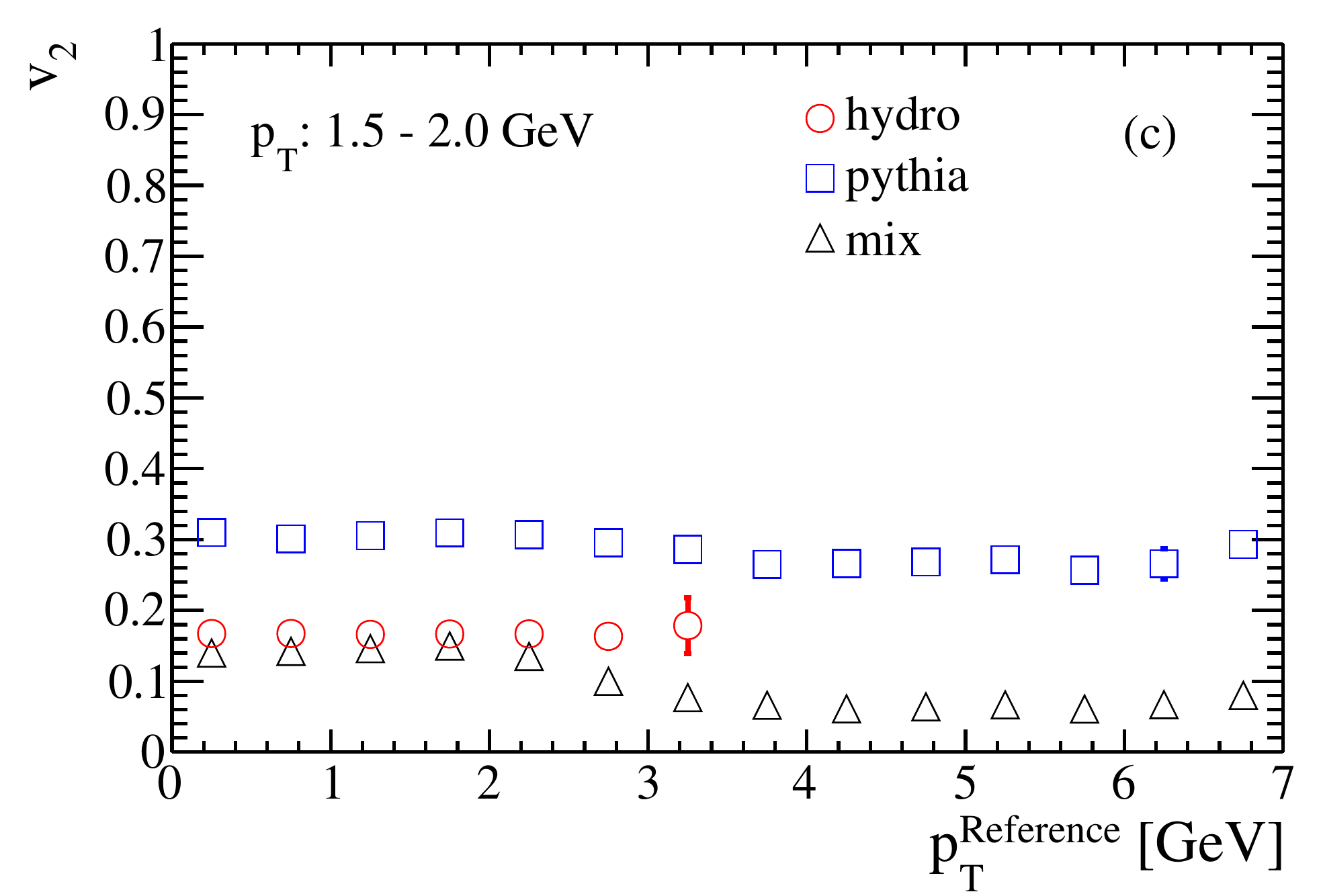} &
\includegraphics[width=0.45\textwidth]{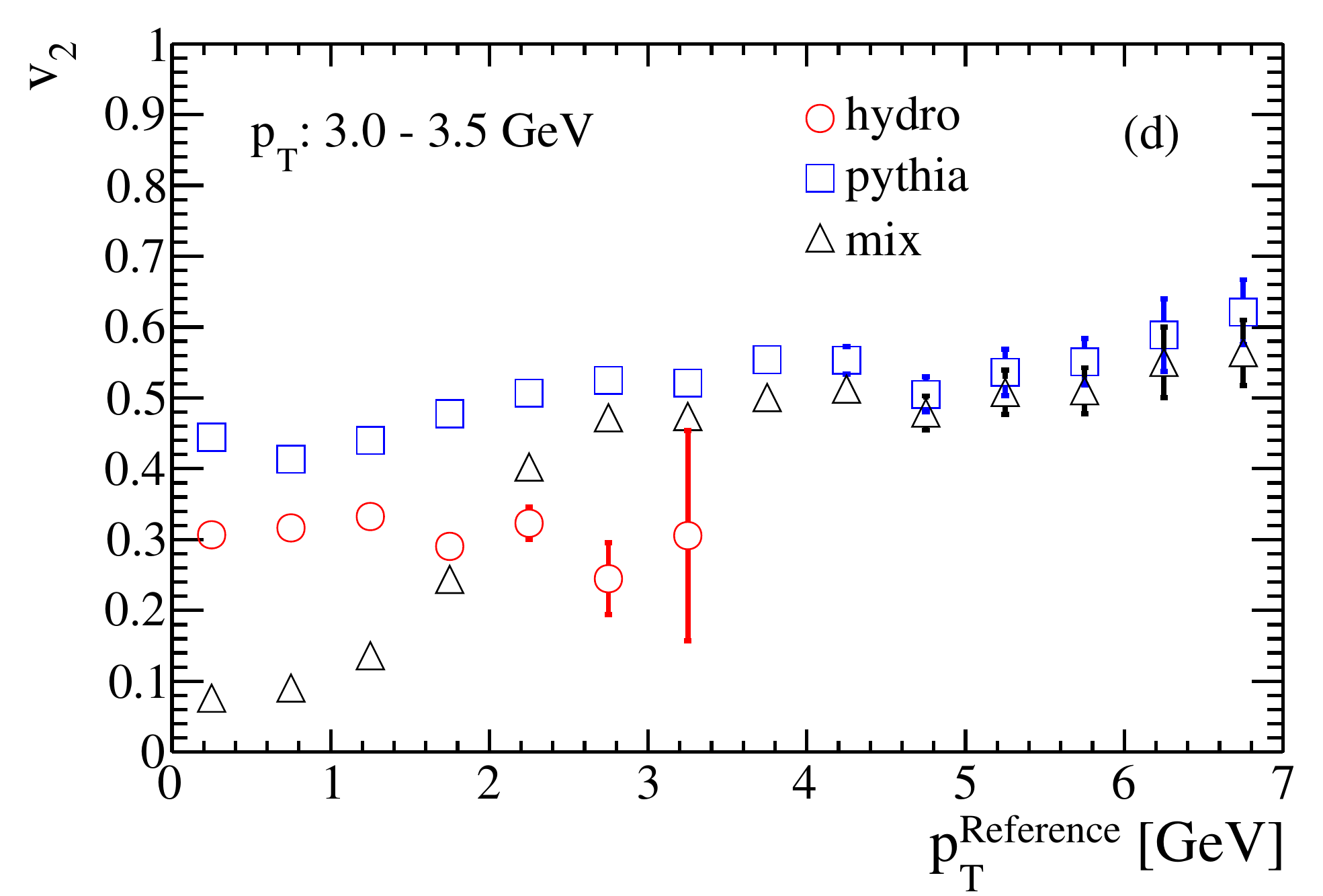} \\
\end{tabular}
\caption{\label{Fig:V2vsPtRefMixed}\label{Fig:pTspectrum:v2VsPtMixed}(Color online) (a) \pt\ spectrum of flow (``hydro'') and nonflow (PYTHIA \pp) event components. (b) $\vTwoDiag(p_T)$ for pure hydrodynamical flow, PYTHIA \pp\ and mixture of both. (c) and (d): $v_{2} = \Vdtwo(p_T,p_T^{\rm{Ref}})/\vTwoDiag(p_T^{Ref})$ as a function of \pTRef\ for pure hydrodynamical flow, PYTHIA \pp\ and mixture of both components for two bin of \pt\ of interest: $1.5 < p_T< 2$ GeV/c (c) and $3 < p_T < 3.5$ GeV/c (d). } 
\end{center}
\end{figure}

\begin{figure}[htdp]
\begin{center}
\begin{tabular}{cc}
\includegraphics[width=0.45\textwidth]{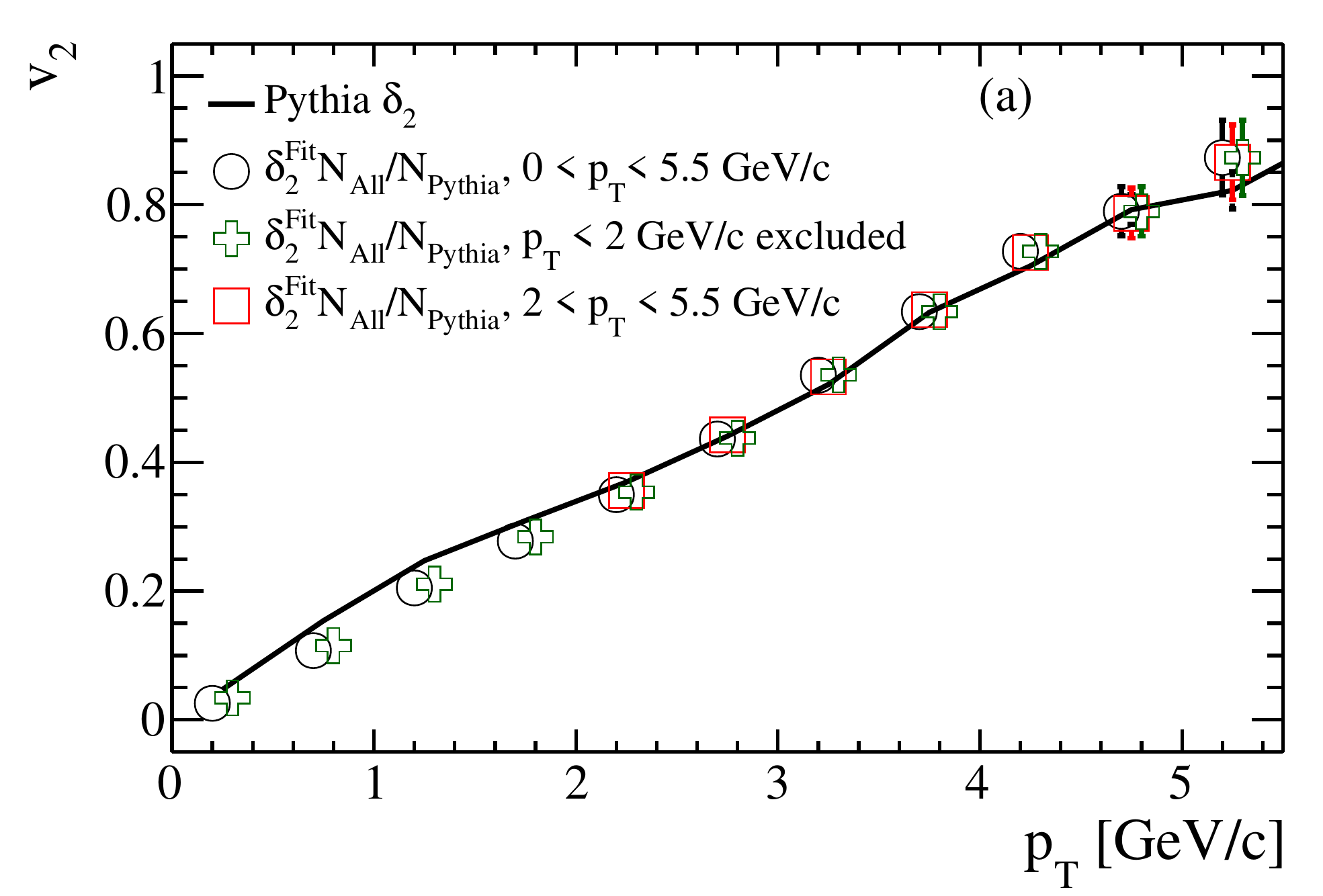} &
\includegraphics[width=0.45\textwidth]{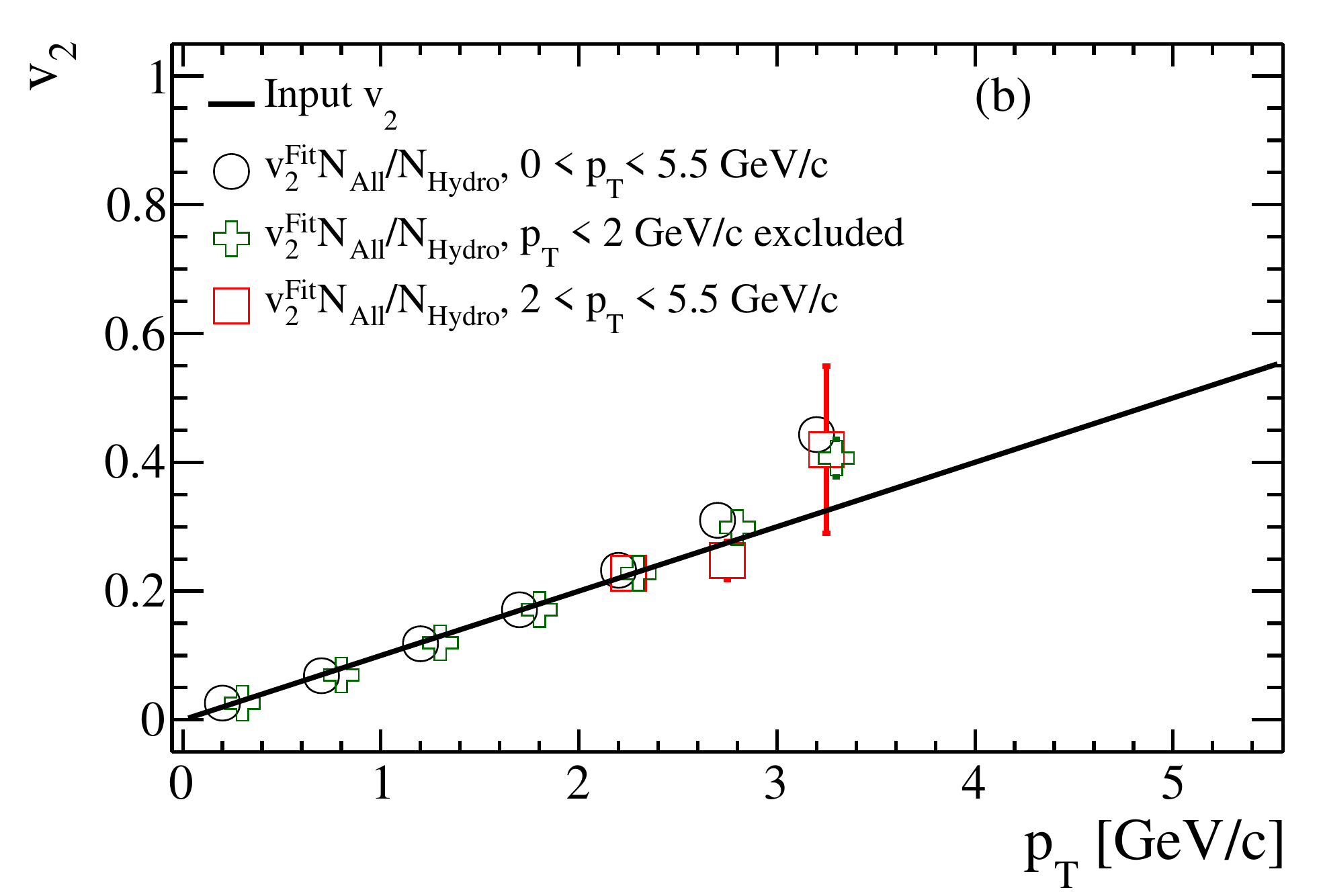} \\
\includegraphics[width=0.45\textwidth]{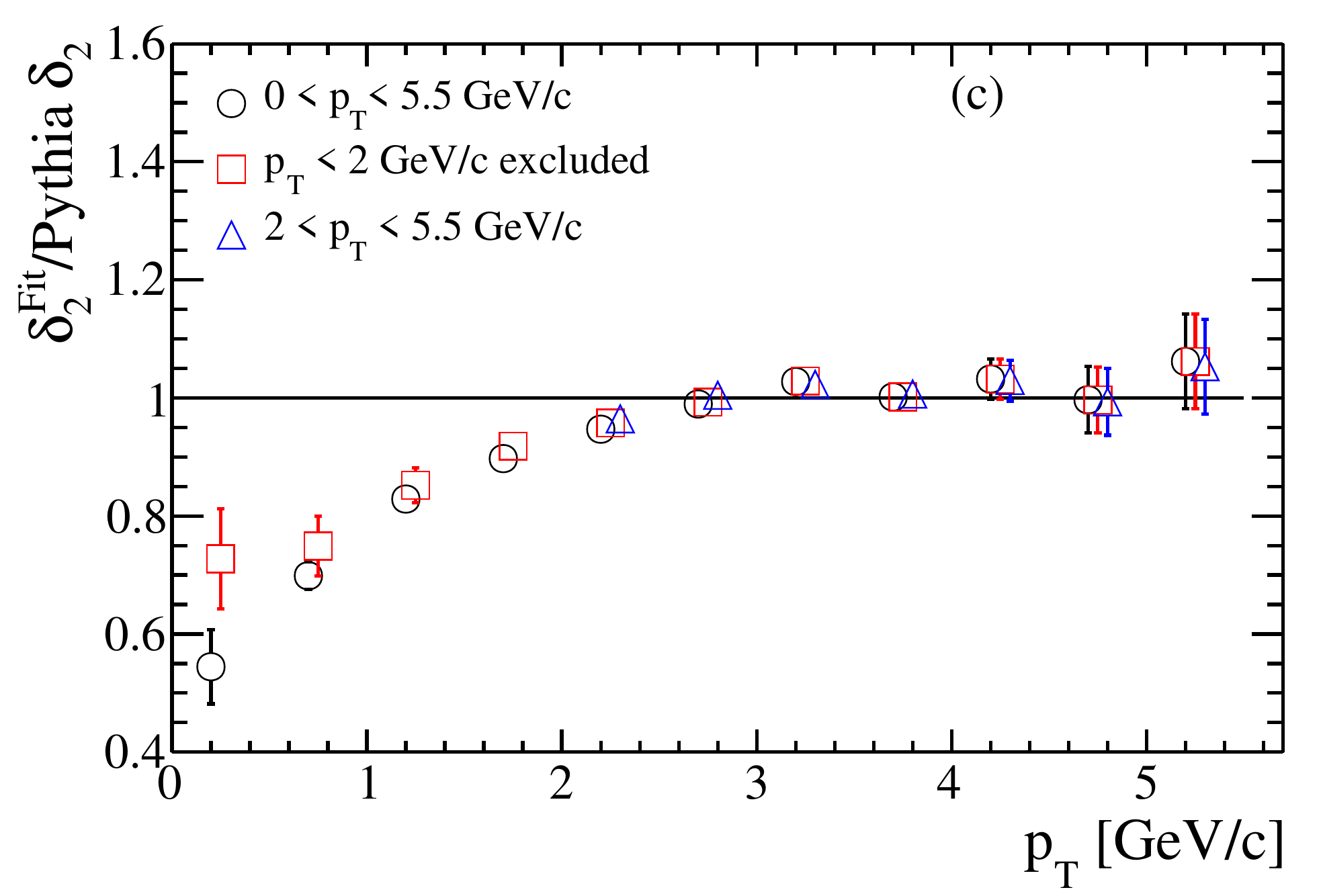} &
\includegraphics[width=0.45\textwidth]{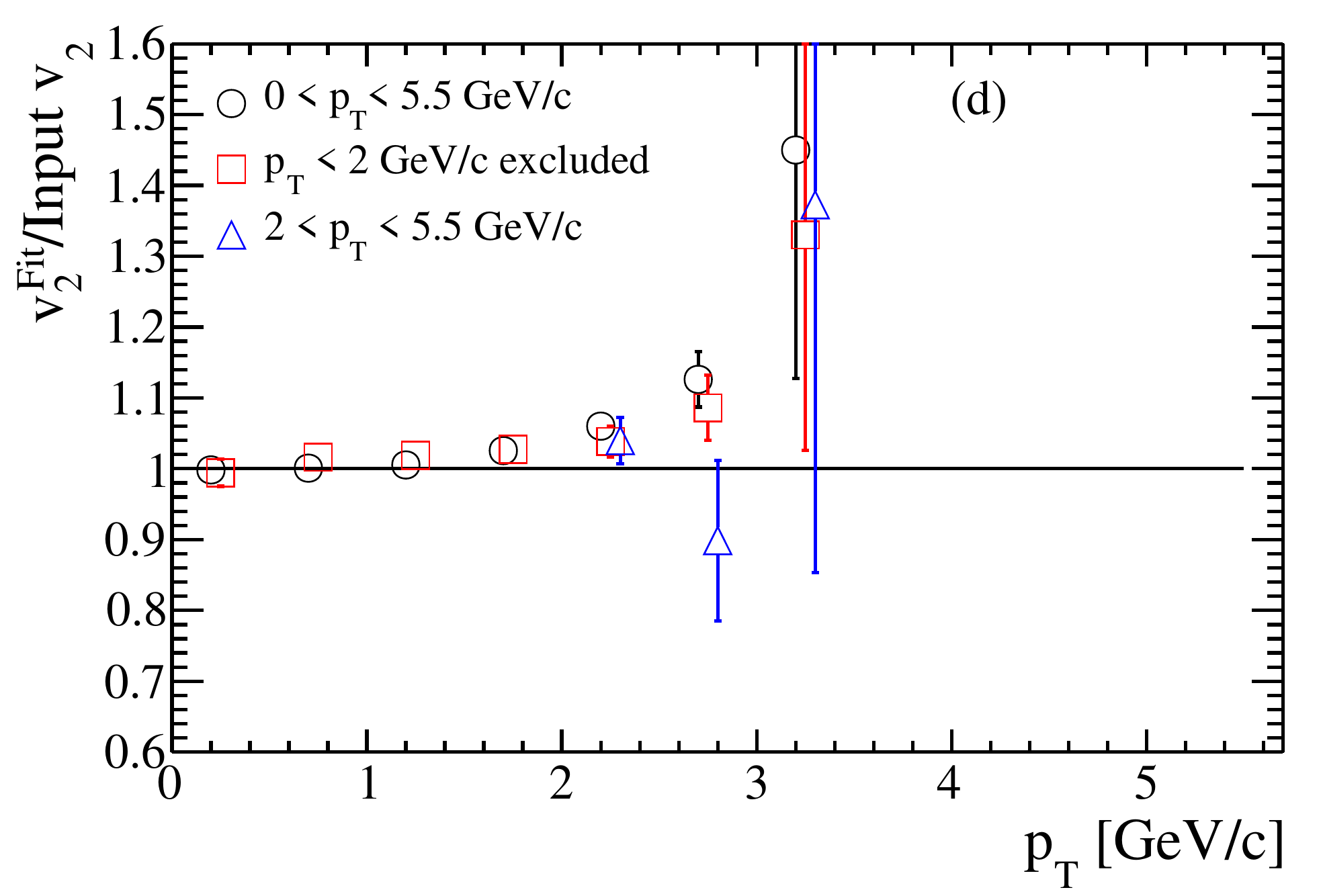} \\
\end{tabular}
\caption{\label{Fig:FitVsExpectations}(Color online) Fitted flow \vn\ and nonflow $\dTwo$ as a function of \pt\ compared to the expected input flow and nonflow (see text for details).} 
\end{center}
\end{figure}

If nonflow factorizes, then Eq. (\ref{Eq:FlowNonFlowFactorization}) can be used in a two component fit to separate flow and nonflow. Note that the separation requires the \pt\ dependence of flow and nonflow to be different, which should be the case in general. In order to study the possibility of separating flow and nonflow quantitatively, we consider a two-component model: hydrodynamical elliptic flow of bulk particles and nonflow (jet) correlations. The former component, which we refer to as ``hydro'', is approximated by a toy model simulation where we generate particles with $v_{\rm{2}} = a \cdot p_{\rm{T}}$, where $a = 0.1$~GeV$^{-1}$, which represents approximately the measured charged hadron elliptic flow for $p_T<2$ GeV/c in mid-central (30 -- 40\%) Au+Au events at \sNN~\cite{Phenix:ChHadFlow:Adare:2010ux}. Those particles are generated from soft \pt\ spectrum $dN/dp_{\rm{T}} \propto p_T \exp{^{-p_{\rm{T}}/B}}$ where $B = 0.23$~GeV/c is obtained from fit to pion \pt\ spectrum at mid-rapidity reported by PHENIX in mid-central Au+Au events at \sNN\ (30--40\% central events)~\cite{InputSpectra:Pi}. For each event a random number of particles from the range of 46 - 133 is generated independently from a uniform distribution in two pseudorapidity regions: $-1 < \eta<-0.5$ and $0.5 <\eta<1$. The multiplicity of generated Monte Carlo tracks corresponds approximately to the uncorrected multiplicity of mid-central Au+Au events at \sNN\ registered in the STAR Time Projection Chamber~\cite{STAR:AuAuCentDef}.  

We mix one PYTHIA event (nonflow correlation) with one ``hydro'' event. The \pt\ spectra of hydro and PYTHIA particles are shown in Fig. \ref{Fig:pTspectrum:v2VsPtMixed}~(a). The PYTHIA event is uncorrelated with the reaction plane of the hydro part of our events for this study. The ``mixed'' events are analyzed with the direct Q-cumulants method.

Figure \ref{Fig:pTspectrum:v2VsPtMixed} (b) shows Fourier coefficient $\vTwoDiag$ as a function of \pt\ calculated for pure ``hydro'', PYTHIA and ``mixed'' events. The results are correlated with relative multiplicities of those components in an event: at low-\pt, where  ``hydro'' particles dominate,  \vn\ follows closely the hydrodynamical flow while at high-\pt, where the jet component dominates, \vn\ is determined by nonflow correlations. Figure \ref{Fig:V2vsPtRefMixed} (c-d) show $v_{2}$ as a function of \pTRef\ for two regions of \pt: (c) $1.5 - 2$ GeV/c, where hydro component dominates, (d) $3 - 3.5$ GeV/c, where nonflow (jets) dominates. The observed dependence on $p_{\rm{T}}^{\rm{Ref}}$ reflects the relative contribution from flow and nonflow components: at low-\pt\ of interest, results are dominated by flow and \vn\ factorizes (is independent on \pTRef) as expected. At higher \pTRef\ \vn\ deviates from ``flow-only" trend and then reassembles the nonflow correlations diluted slightly by hydro particles (see Fig. \ref{Fig:V2vsPtRefMixed} (c)). However, \vn\ seems to factorize in the nonflow region as well (Fig. \ref{Fig:V2vsPtRefMixed} (c-d)). Note that in our mixed events the PYTHIA event is not correlated with the reaction plane of the hydro event. Adding in a correlation would change the details of the \pTRef\ dependences in Fig. \ref{Fig:V2vsPtRefMixed} (c-d). Moreover, even though the hydrodynamic flow dominates at low-\pt, it is slightly diluted by low-\pt\ particles from the embedded \pp\ event. The contribution of hydro flow to the measured $v_{2}$ scales with the ratio of number of hydro particles to the overall number of particles: $v_{2}(p_T) = v_{2,hy}(p_T) N_{\rm{hy}}(p_T)/[N_{\rm{hy}}(p_T) + N_{\rm{PYTHIA}}(p_T)]$, as in Eq. (\ref{Eq:Flow:ClusterModel}) (where $v_{2,\acl} = 0$ because our embedded PYTHIA event is random to the reaction plane). Similarly, the measured nonflow is given by $\delta_{2}(p_T) = \delta_{2}^{\rm{PYTHIA}}(p_T) N_{\rm{PYTHIA}}(p_T)/[N_{\rm{hy}}(p_T) + N_{\rm{PYTHIA}}(p_T)]$ as in Eq. (\ref{Eq:NonFlow:ClusterModel}) where $\Ncl = 1$ in our case.

We now fit formula (\ref{Eq:FlowNonFlowFactorization}) to the $\Vdtwo$ data of the mixed events. Fits were limited to  $p_T < 5.5$ GeV/c for ``mixed'' events to ensure sufficient statistical precision of the data. We expect that nonflow factorization works best at high-\pt. Therefore we compared fits in three different cases: (a) $0<p_T^a <5.5$ GeV/c and $0<p_T^b<5.5$ GeV/c, (b) $2<p_T^a<5.5$ GeV/c or $2<p_T^b<5.5$ GeV/c i.e. points with low \pt\ pairs are excluded from fit, (c) $2<p_T^a<5.5$ GeV/c and $2<p_T^b<5.5$ GeV/c i.e. only high-\pt\ points are used in the fit. In each case fit quality is good: $\chi^2/NDF \approx 1$. Goodness of the fit at low-\pt\ for mixed events is better than in the case of \pp\ data alone because low-\pt\ is dominated by hydro flow in the mixed events, which factorizes.  In order to compare the fit results with expected values of \vn\ and $\delta_2$, we present the ratio of \vn\ and $\dd{2}$, scaled by relative multiplicity, to the input parameters (Fig. \ref{Fig:FitVsExpectations} (c) and (d)). In the case of non-flow, there is a $30 - 45\%$ deviation at lowest-\pt; at $p_T > 2$~GeV/c good agreement is observed. This is expected because we showed in Sec. \ref{Sec:Nonflow:factorization} that nonflow factorization works best at high-\pt. The fit result for flow is also good - there is a small deviation (less than a few \%) in the region where data provides sufficiency statistical precision ($p_T < 3$ GeV/c).

We note that the fit of a quadratic function of the form of Eq. (\ref{Eq:FlowNonFlowFactorization}) to the $\Vdtwo$ data is a mathematical problem with possibly multiple solutions. Different solutions can result from a different set of initial values of the fit parameters. However the solution representing the real physical correlations can be distinguished from the purely mathematical ones when the high \pt\ region is considered, where the nonflow correlations dominate and anisotropic flow is expected to be small.

It is worthwhile to examine quantitatively the higher harmonic anisotropic flow and two-particle azimuthal correlations measured in Pb+Pb collisions \cite{ALICE:Higher:Harmonics,Alice:FlowFactorization, ATLAS-CONF-2011-074}. In Ref. \cite{ALICE:Higher:Harmonics} the higher harmonic $v_n\{2\}$ (n = $2,3,4,5$) were first obtained from two particle correlations, where particles of interest \textit{a} with a given $p_T^a$ were correlated with reference particles \textit{c} of all \pt. The azimuthal correlations between triggered and associated particles were then compared to correlation shapes expected from the measured $v_2\{2\}$, $v_3\{2\}$, $v_4\{2\}$, $v_5\{2\}$ harmonics evaluated at the \pt\ corresponding to trigger and associate particles.   It was shown that the measured Fourier coefficients describe the trigger-associate correlations. Let us examine what difference would be expected between the trigger-associate correlations and the composition of products of the Fourier harmonics measured with a third particle \textit{c}. In the case of two-component factorization, $v_n\{2\}$ for the trigger particle \textit{b} is given by $v_n'(p_T^b) = (v_{n}(p_T^b)v_{n}(p_T^c) + \delta_n(p_T^b) \delta_n(p_T^c) ) /\sqrt{v^2_{n}(p_T^c) + \delta^2_n(p_T^c)}$, where the denominator is the \textit{c} particle harmonic measured by two-particle correlation with both particles from all \pt. Likewise for the anisotropic flow $v_n'(p_T^a)$ of associate particle \textit{a}. The deviation of trigger-associate correlation from composition of single particle harmonics would be 
\begin{multline}
\frac{\Vdn(p_T^b,p_T^a)}{v'_n(p_T^a) v'_n(p_T^b)} -1 =  \frac{v_{n}(p_T^b)v_{n}(p_T^c) + \delta_n(p_T^b) \delta_n(p_T^c)}{\frac{v_{n}(p_T^a)v_{n}(p_T^c) + \delta_n(p_T^a) \delta_n(p_T^c)}{\sqrt{v^2_{n}(p_T^c) + \delta^2_n(p_T^c)}} \frac{v_{n}(p_T^b)v_{n}(p_T^c) + \delta_n(p_T^b) \delta_n(p_T^c)}{\sqrt{v^2_{n}(p_T^c) + \delta^2_n(p_T^c)}}  } -1 \\
\approx \left( \frac{\delta_n(p_T^a)}{v_n(p_T^a)} - \frac{\delta_n(p_T^c)}{v_n(p_T^c)} \right) 
\left( \frac{\delta_n(p_T^b)}{v_n(p_T^b)} - \frac{\delta_n(p_T^c)}{v_n(p_T^c)} \right) 
\label{Eq:FactTriggered}
\end{multline}
in case of small nonflow relative to flow. Another  approach is to take the trigger (and associate) particle harmonic directly from two-particle correlations in the same trigger (associate) \pt\ region \citep{ATLAS-CONF-2011-074}. The deviation in this approach would be
\begin{equation}
\frac{\Vdn(p_T^b,p_T^a)}{v'_n(p_T^a) v'_n(p_T^b)} -1 =
\frac{\Vdn(p_T^b,p_T^a)}{\sqrt{\Vdn(p_T^a,p_T^a)}\sqrt{\Vdn(p_T^b,p_T^b)}} -1
\approx - \frac{1}{2} \left( \frac{\delta_n(p_T^b)}{v_n(p_T^b)} - \frac{\delta_n(p_T^a)}{v_n(p_T^a)} \right)^2 
\label{Eq:DevFromFactorization}
\end{equation}

As seen from Eqs. (\ref{Eq:FactTriggered}) and (\ref{Eq:DevFromFactorization}), even for relatively large nonflow $\delta_n(p_T)/v_n(p_T) \approx 10\%$, the deviation amounts to only the order of 10$^{-3}$ and would be very challenging to observe with current statistical precision of the data, which is roughly what the results in Refs. \cite{ALICE:Higher:Harmonics,ATLAS-CONF-2011-074} indicate. Therefore it is not surprising that global fit in Ref. \cite{Alice:FlowFactorization} gave satisfactory results at low-\pt. If $\delta_n(p^a_T)/v_n(p^a_T) =\delta_{n}(p^b_T)/v_n(p^b_T)$ i.e. flow and nonflow have the same \pt\ dependence, then $\Vdn(p^a_T,p^b_T)$ factorizes precisely even though nonflow is present. 

\section{Summary}

We have shown by PYTHIA studies of \pp\ collisions that nonflow correlations due to jet fragmentation approximately factorize.
Based on this observation we developed a novel method of fitting a two-component factorization expression to heavy ion data in order to extract anisotropic flow and nonflow. We used PYTHIA events embedded in toy-model Monte Carlo simulations of hydrodynamical flow to demonstrate that this method works well for a mixture of flow and nonflow.

We focused on the second harmonic (\vn\ and $\delta_2$) in this paper. In principle, a two-component factorization fit should work for other harmonics as well, although the sign of nonflow contribution has to be established for specific case. For instance, jet correlation from PYTHIA would give a negative contribution and therefore $V_{3 \Delta}(p_T^a,p_T^b) = v_{3,hy}(p_T^a)v_{3,hy}(p_T^b) -  \delta_3(p_T^a) \delta_3(p_T^b)$ should be used in the fit.

\acknowledgments{}

This work was supported by US Department of Energy under Grant
DE-FG02-88ER40412 and Japanese Ministry of Education, Culture, Sports, Science, 
and Technology and the Japan Society for the Promotion 
of Science.

\bibliography{bibligraphy}{}
\bibliographystyle{apsrev}

\end{document}